# Light management in highly-textured perovskite solar cells: From full-device ellipsometry characterization to optical modelling for quantum efficiency optimization

Chenxi Ma[1], Daming Zheng[2], Dominique Demaille[1], Bruno Gallas[1], Catherine Schwob[1], Thierry Pauporté[2] and Laurent Coolen[1]

[1] *Sorbonne Université, CNRS, Institut de NanoSciences de Paris, INSP, F-75005 Paris, France,*

[2] *Chimie ParisTech, PSL Research University, CNRS, Institut de Recherche de Chimie Paris (IRCP), UMR8247, 11 rue P. et M. Curie, F-75005 Paris, France.*

**Abstract :**

While perovskite solar cells (PSCs) are now reaching high power conversion efficiencies (PCEs), further performance improvement requires a fine management and an optimization of the light pathway and harvesting in the cells. These go through an accurate understanding, characterization and modelling of the optical processes occurring in these complex, often textured, multi-layered systems. In the present work, we have considered a typical methylammonium lead iodide (MAPI) solar cell built on a fluorine-doped tin oxide (FTO) electrode of high roughness (43 nm RMS). By variable-angle spectroscopic ellipsometry (VASE) of the full PSC device, we have been able to determine the optical constants of all the device layers. We have designed a one-dimensional (1D) optical model of the stacked layers where the rough texture is described as layers of effective-medium index. We have supported the model using data extracted from scanning electron microscopy, diffuse spectroscopy and photovoltaic efficiency measurements. We show that the 1D model, while insufficient to describe scattering by the FTO plate alone, gives an accurate description of the full device optical properties. By comparison with the experimental external quantum efficiency (EQE), we estimate the internal quantum efficiency (IQE) and the effect of the losses related to electron transfer. Based on this work, we finally discuss the optical losses mechanisms and the possible strategies that can be implemented to improve light management within PSC devices and further increase their performances.

**Introduction :**

Hybrid organic-inorganic perovskite solar cells (PSC) have rapidly emerged as a promising candidate for efficient and cost-effective photovoltaic systems due to their strong light absorption, high charge carrier mobility and low processing costs.[1] Recent developments of PSCs include the use of interfacial SAMs[2] or 2D layers,[3] multiple cations perovskites[4] or the use of additives that control the layer composition and crystallinity.[5] This technology reaches a present record efficiency of 25.5%.[6] Commercialization of single junction PSC or in tandem with a silicon solar cell is under study by various industrial companies.[7] As perovskite materials now attain high performances and a rather good stability, further advances can be expected by optimizing light penetration in the devices and harvesting by the perovskite absorber.[1,8-22] Accurate optical models are necessary in order to explore such light management strategies, as well as to distinguish optical effects from charge separation and transfer effects in the overall experimental photovoltaic efficiency. The final target is to optimize the light management to further boost the cells performances. It requires a precise knowledge of the optical properties (index, thickness, roughness) of each layer of the device.

Ellipsometry is one of the best techniques for probing the optical properties of stacked thin films. It has been used in many experimental reports on PSC to determine the optical indices of each device material.[17-27] In silicon or organic photovoltaics, ellipsometric characterization of a full device stacking has also been discussed in a perspective of monitoring industrial fabrication processes.[28] However, data treatment and interpretation can become problematic when stacks of many layers must be considered,[29] as in the case for a full solar cell device. For this reason, reports on optical modelling of PSCs usually use either literature indices or ellipsometry characteristics of the material obtained independently as a single layer on a standard substrate.[13,18-23, 30] Few works have reported ellipsometry investigations on full PSC devices to validate the optical model,[20,31] even though materials deposition is known to depend critically on the substrate.

Modelling the device as a stack of independently-characterized layers is all the more difficult and possibly inappropriate in the presence of rough interfaces. The layer of transparent conducting oxide employed as the front contact can present a rough texture[16-18,28,32] of up to hundreds of nanometers in height[30] with possible effects on the device's efficiency.[8] Some deposition protocols for the perovskite and other layers can also generate a significant roughness.[9, 13, 33] Rough interfaces affect solar cell optical properties and intentional texturing can be a tool for light management because (i) it creates an effective index gradient which facilitates light penetration into the active absorber layer and (ii) it scatters light to higher internal propagation angles so that it remains trapped within the device and is better absorbed.[1, 8-12, 34] Index-gradient effects can be probed by ellipsometry, from which a one-dimensional (1D) model of stacked planar layers can be derived and treated by the transfer-matrix method (TMM). Rough interfaces are then treated by introducing intermediate planar layers of mixed composition described by an effective medium approximation (EMA).[29] On the other hand, light scattering cannot be quantified by specular ellipsometry nor described by a 1D model of planar layers. Diffuse spectroscopic methods and 3D simulations must then be used.

It is thus necessary to establish accurate ellipsometry protocols for the full PSC devices characterizations, particularly in the case of highly textured interfaces, and to assess the validity of the derived optical 1D models by in-depth comparison with other available experimental data and 3D simulations. In this paper, we investigate a typical PSC structure built on a front electrode with a very rough surface. We characterize the device by ellipsometry at each step of the deposition, leading to a model of the structure as a 1D stack of planar layers, with EMA layers describing textured interfaces. We then analyze the validity of this 1D model as compared to 3D numerical simulations by confronting them with other experimental data obtained by electron and atomic-force microscopy, direct and diffuse transmission spectroscopy. From the device's optical behavior determined theoretically, we have then been able to determine the theoretical light harvesting efficiency spectrum that we have compared with the experimental external quantum efficiency (EQE) of the cell. Based on this analysis, we finally discuss some potential light-management improvement paths.

**Results and discussion :**

For this analysis, we chose the typical perovskite solar cell stacking presented in Fig. 1(a). The cells, fully detailed in Ref. [35], were based on methyl ammonium lead iodide ($CH_3NH_3PbI_3$, MAPI) and reached a power conversion efficiency up to 19%. A $SnO_2$:F (fluorine-doped tin oxide - FTO) layer was used as the front contact electrode. These commercial substrates (TEC7, Pilkington - hereafter named "FTO plate") are made of a soda-lime glass slide, covered by a thin $SnO_2$ layer, a thin $SiO_2$ layer, and the FTO layer.[18,28] The role of the two oxide sub-layers is to avoid color effects due to interferences.[28] A compact $TiO_2$ layer was deposited by spray-pyrolysis on top of the FTO as the electron transporting layer (ETL), followed by a mesoporous $TiO_2$ layer (30-nm $TiO_2$ nanospheres) in order to maximize the charge collection and transfer to the ETL.[36] The active MAPI layer was then spin-coated as described in ref. 37 : Figure 1(c) shows a SEM image of the sample cross-section at this stage. It was then covered by a layer of 2,2',7,7'-tetrakis(N,N'-di-p-methoxyphenylamine)-9,9'-spirobifluorene (Spiro-OMeTAD) which acts as the hole transporting layer (HTL) and by a gold film acting as the back-contact electrode.

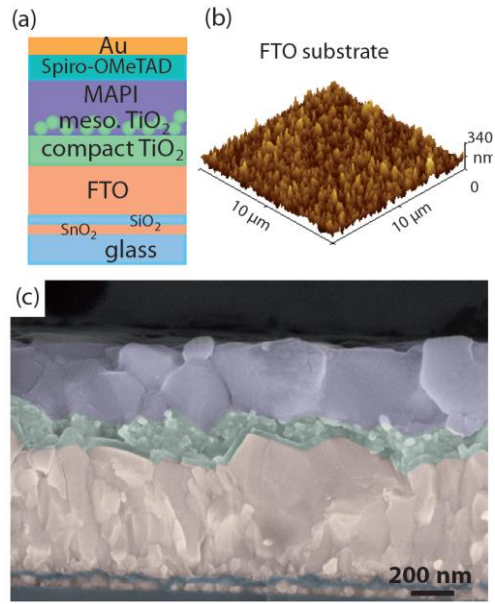

*Figure 1 : (a) Scheme of the solar cell structure. (b) Atomic-force microscopy (AFM) image of the surface of the FTO layer. (c) Scanning electron microscopy (SEM) cross-sectional view of the sample after MAPI deposition (before Spiro-OMeTAD and gold deposition ; the colors are added as a guide for the eye).*

FTO is the most popular transparent conducting oxide employed in PSCs.[28] As compared to indium tin oxide (ITO), it offers lower costs and a better stability and resistance to high temperatures. However, many manufactured FTO samples are known to present high surface roughness (RMS).[8, 17, 28, 30] From AFM images of the FTO surface (fig. 1(b)), we found in our case a root-mean-square roughness of 43 nm. The SEM profile (fig. 1(c)) also shows clearly the FTO texture, with peak-to-peak heights as large as 100-200 nm. In spite of the large FTO roughness, the two $TiO_2$ layers, compact and mesoporous, were well-covering and followed almost perfectly the underlying FTO texture profile. As a result, the roughness of these layers, measured by AFM at 39 and 27 nm, respectively, were close to the roughness of the FTO surface. On the other hand, the MAPI layer was much thicker and smoothed the surface (Fig. 1(c)).

Ellipsometry relies on measuring the ratio of the p-polarized to s-polarized reflection spectra and fitting them with a one-dimensional (1D) model of planar layers. The fitting parameters are thus the index and thickness of each layer. The roughness of an interface between two materials is described in the model by adding one or several layers containing a mixture of the two materials. The index of the mixture was described here within a Bruggeman-type effective medium approximation (EMA) and the fitting parameters were the EMA layer thickness and the composition of the mixture. When considering a stack of numerous layers, including some large roughness, many variables must thus be adjusted to fit the experimental data and it is necessary to support the resulting optical model by as much experimental data as possible.[29] In this study, we performed variable-angle spectroscopic ellipsometry (VASE) at three different incidence angles, and at each step of the device deposition. Moreover, preliminary estimates for the indices of the materials were either determined independently by depositing each material individually on a glass substrate and performing an ellipsometric characterization, or were found in literature tables for the most standard materials (see Supporting information for the protocol and resulting index curves). All optical constants extracted from ellipsometry were found in excellent agreement with literature data. The mesoporous $TiO_2$ layer was well-described by an EMA layer of air and $TiO_2$, as expected since the $TiO_2$ spheres were much smaller (~30 nm diameter) than the optical wavelength.

Figure 2 displays the measured ellipsometric data (full lines) for the FTO commercial substrate (Fig. 2(a)), for each new deposited layer (Figs. 2(b) to (e)) and for the full device (Fig. 2(f)). At each step (a) to (f), a 1D optical model of the system was established (right column of fig. 2) yielding theoretical curves (dotted lines) which were all in very good agreement with the experimental spectra. The FTO-$TiO_2$-MAPI portion is quite complex since the amplitude of the FTO texture is larger than the thickness of the two upper $TiO_2$ layers. Therefore a ternary mixture of FTO, $TiO_2$ and MAPI had to be considered, and three successive EMA layers were used. No adjustment of the pre-determined indices was necessary in order to optimize the model agreement with the data. The only free parameters were the thickness of each layer and the composition of the three EMA mixtures. At each new deposition step, unless stated otherwise in the figure, a good fit could be obtained while keeping the same model for the underlayers and just replacing air by the newly-deposited material and adding more layers on top. From Fig. 2(e) to (f), a significant change of the Spiro-OMeTAD thickness had to be introduced, from 192 to 228 nm, but Figure 2(f) is dominated by the Au absorption so that the Spiro-OMeTAD thickness might be less precise than the value from Fig. 2(e). The thicknesses obtained from ellipsometry well-match those estimated from the SEM image of Fig. 1(c). The index profile within the stack of three EMA layers is in acceptable agreement with the index profile calculated by averaging over the 3D structure as known from the SEM image (see detailed discussion in section B of the Supporting information). Apart from this portion, no other EMA layer needed to be introduced to describe the roughness of the other interfaces.

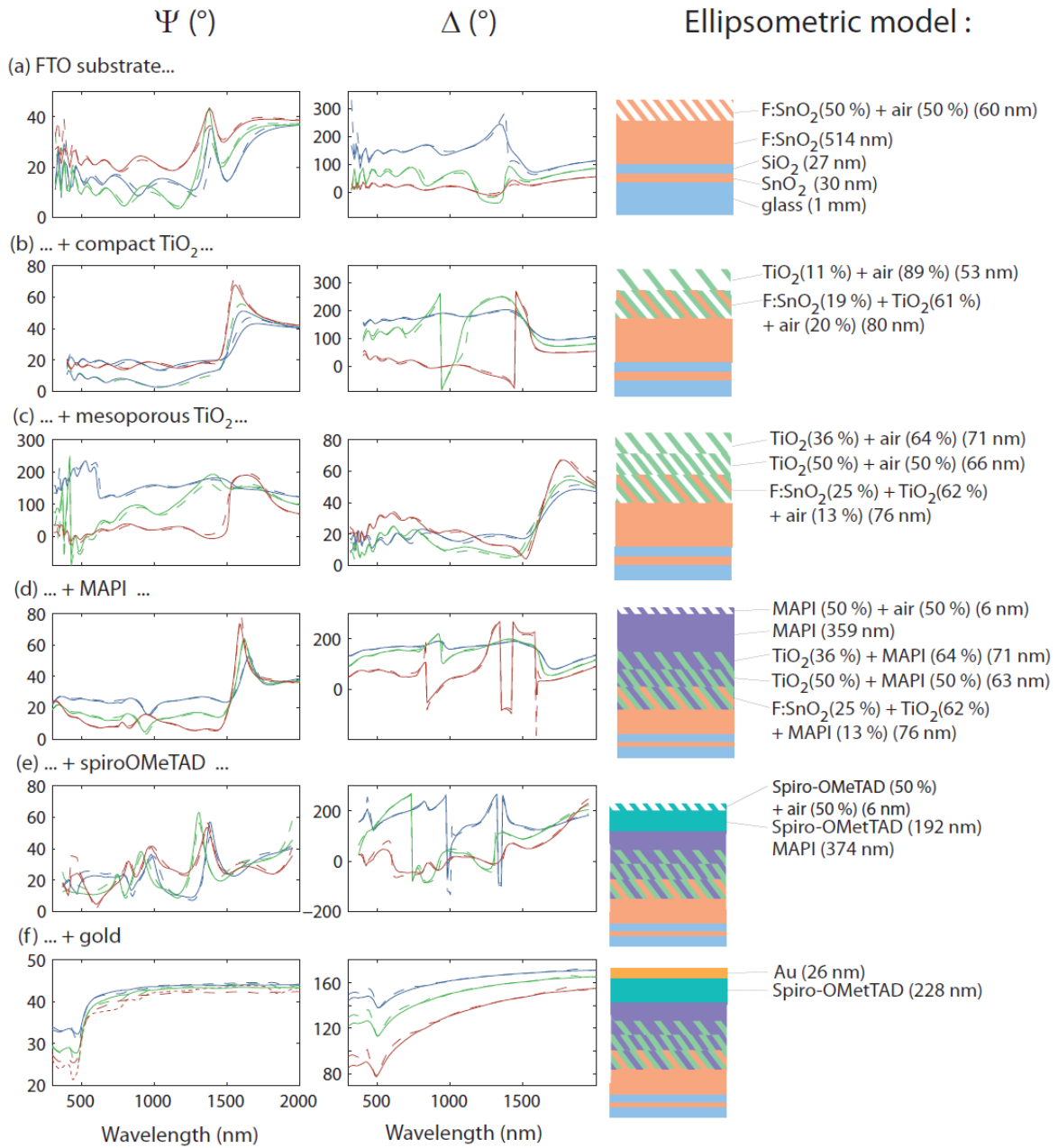

*Figure 2 : Left and middle columns : measured (full lines) and fitted (dotted lines) ellipsometric curves at each successive step of sample deposition, at incidence angles 50° (blue), 60° (green) and 70° (red). Right column : model structure obtained by fitting the ellipsometric data (EMA layers are represented by hatches). This structure will define the 1D model used in the following of the paper.*

Although ellipsometry data are consistent with the 1D model at each step of the device deposition, some other optical properties are absolutely not well-described by this model. In particular, Fig. 3(a) compares the measured total transmission spectrum of the FTO plate with the spectrum calculated according to the 1D model of this FTO plate (shown in Fig. 2(a)). The two curves differ significantly over the whole visible range, in particular at blue wavelengths. Scattering by the rough sample surface is the likely explanation for this difference, as it is anticipated to decrease the transmitted intensity and have a stronger effect at lower wavelengths (wavelengths which are comparable to the 100-200 nm scale of the roughness patterns). This is confirmed by measuring the diffuse transmission spectrum with an integrating sphere (Fig. 3(b)). A significant fraction of the total light incident intensity is scattered by the FTO layer: up to 20 % at 350 nm. While the effective index gradient induced by the roughness can be well-probed by ellipsometry, as demonstrated in Fig. 2, scattering is not well-analyzed because ellipsometry relies on the ratio between s- and p-polarizations so that it is little affected by reflection losses due to scattering (as long as they have low polarization dependence).

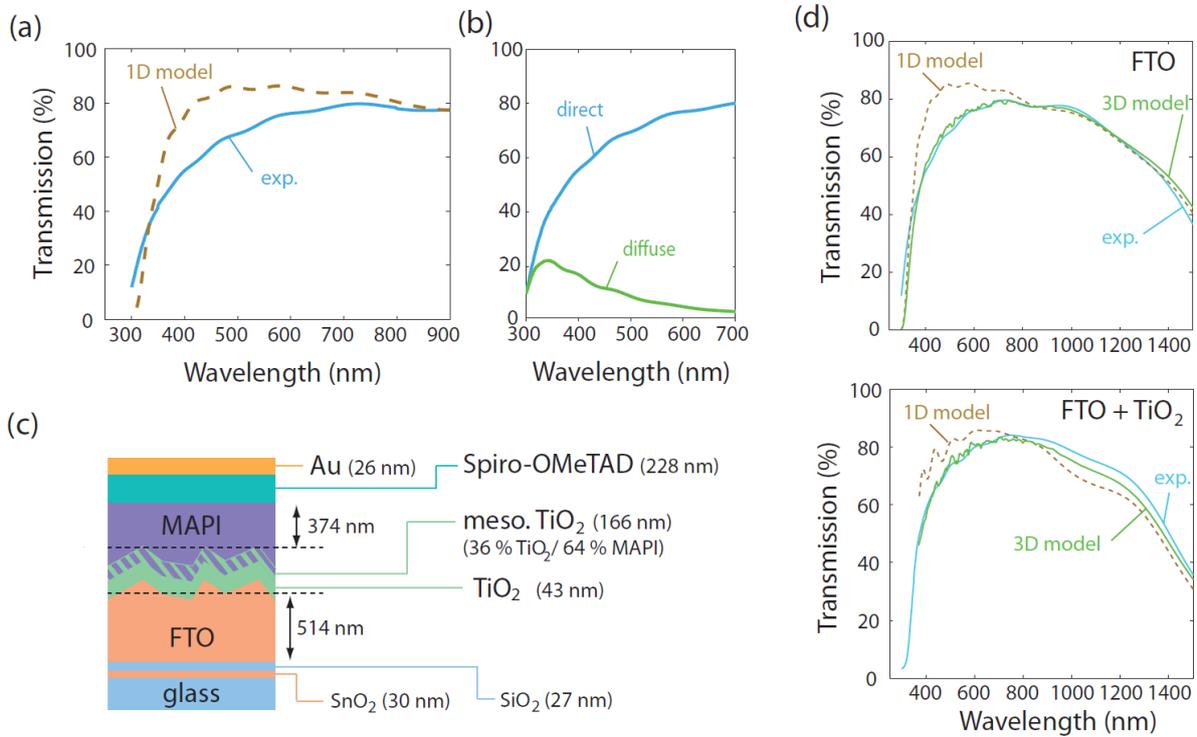

*Figure 3 : (a) Direct transmission spectrum of the FTO substrate : experimental (full line) and simulated within the 1D model of fig. 2(a) (dotted line). (b) Direct (blue) and diffuse (green) transmission spectra of the FTO substrate at normal incidence. (c) Schematic of the full-stack 3D model structure for the finite-elements simulation. The FTO upper surface has a random profile of RMS roughness 50 nm. (d) Transmission curves of the FTO plate and the FTO plate covered by a compact $TiO_2$ layer : experimental (blue line), calculated from the 1D model described in figs. 2(a) and (b) (yellow dotted line), modelled by the 3D finite-element calculation with a rough surface (green line).*

Since 1D model, by definition, can only describe direct transmission and reflection, it is necessary to build a numerical 3D model of the sample in order to well-describe scattering effects. We designed our 3D model as depicted in Fig. 3(c) for the full device ; for the FTO plate alone and for the intermediate deposition steps, the same model was used with the upper layers replaced by air. We considered for all materials the same thicknesses and indices as determined by ellipsometry (Fig. 2(f)). Mesoporous $TiO_2$ was described as an EMA layer, like for the 1D model : we will not discuss effects of scattering by the $TiO_2$ spheres as their mean diameter is much smaller than any wavelength considered. A random surface texture was introduced for the FTO plate with 50-nm RMS roughness. For the compact and mesoporous $TiO_2$ layers, their texture was a translation of the FTO layer texture (see the structure profile in Fig. 5c), with a reduced RMS roughness (30 nm) for the mesoporous layer in order to take into account its lower roughness measured by AFM. The thickness of the compact and mesoporous $TiO_2$ layers cannot be extracted from the ellipsometric data and was assumed to be the same as when they are deposited independently on glass.

Figure 3d plots the experimental transmission spectra of the FTO plate (left) and of the FTO plate covered by the compact $TiO_2$ layer (right), superimposed with the results of the 1D model and of the 3D model. For both samples, the 1D model overestimates the actual transmission, especially in the visible range, because it does not take into account scattering losses. The 3D model, on the other hand, describes the experimental data with an excellent precision. For the FTO plate, the agreement is better than 1.5 % above 400 nm. For the FTO+TiO$_2$ sample, it is slightly less good due to the increased sample complexity but remains better than 4.5 % (standard deviation 2 %). The 3D model thus provides an excellent modelling of the experimental structures, even when the roughness causes a strong scattering, while the 1D model is not able to describe scattering effects in the transmission spectra.

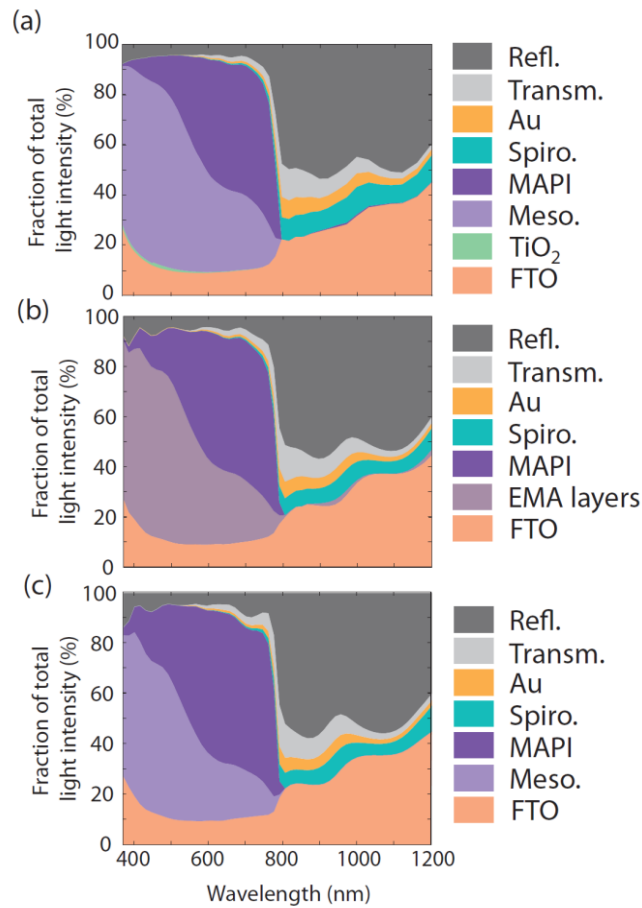

*Figure 4 : Theoretical fractions of light absorbed by each layer of the solar cell and reflected or transmitted by the device, obtained (a) from the 3D model of roughness (as described in fig. 3(c)), (b) from the 1D model of roughness (as described in fig. 2(f)), (c) without including any roughness (planar FTO-TiO$_2$-mesoporous TiO$_2$-MAPI interfaces). Absorption by the compact TiO$_2$ layer is negligible in (b) and (c). For simplicity under "FTO" are included all layers of the FTO plate : SnO$_2$, SiO$_2$ and SnO$_2$:F.*

The relevant optical information for PSC devices is the ratio between perovskite absorption, which generates the charges, on the one hand, and transmission and reflection of the device and parasitic absorption by the other layers, on the other hand. While transmission and reflection can be probed experimentally, the absorption of each layer can only be accessed through simulations. Figure 4 compares the results of the 3D model (Fig. 4a) and of the ellipsometric 1D model (Fig. 4b). Note that in the 3D model the active layers are the MAPI layer and the mesoporous layer (which contains MAPI and TiO$_2$ but TiO$_2$ absorption is negligible above 400 nm). On the other hand, in the 1D model, the active layers are the MAPI layer and the three EMA layers but only part of the latter absorption is active for charge generation since it includes also FTO absorption. Within both models, absorption by the active layers dominates the visible spectrum. The first 150-200 nanometers of the active medium (corresponding to the mesoporous layer in the 3D model and the EMA roughness layers in the 1D model) contribute to most of the absorption in the blue range and of a significant portion in the red range. This can be explained by the strong MAPI absorption: as shown in the electric field distribution (Fig. S4), most of the incoming beam is absorbed already by MAPI in the mesoporous region and does not reach the MAPI capping layer. We have also noted that the absorbance of the mesoporous layer follows the absorbance curve of MAPI (Fig. S5). MAPI absorption is especially strong at lower wavelengths due to the higher imaginary index component.[18-19, 24, 30] Parasitic absorption is caused essentially by FTO, as calculated previously on similar systems,[18] while the front electrode (air-glass interface) causes around 4 % reflection.

The plots of Figs. 4a and 4b are in excellent agreement, with only a few fringes added in fig. 4b to the reflection component because of interferences in the 1D model (which are not present in the 3D model because of the rough interfaces). This shows that the 1D model obtained from our spectroscopic ellipsometry protocol provides an accurate description of the absorption of the various layers of the full PSC structure. As compared to the case of the FTO plate alone (fig. 3), the full device is probably less affected by scattering because in the FTO plate the index contrast at the FTO/air interface is very high, while in the device the index contrast between FTO, $TiO_2$ and MAPI is much lower. We also plot in Fig. 4c the result of a 1D structure where no FTO roughness is included (only planar FTO-$TiO_2$-mesoporous $TiO_2$-MAPI layers). The optical behavior of the planar structure is very similar to the rough sample. This indicates that, even though the sample roughness creates an effective gradient index which might modify the optical properties, this effect is quite low, again because of the low index contrast between FTO, $TiO_2$ and MAPI. Eventually, the similarities between figs. 4a, b, and c show that the large roughness of the FTO surface has no significant effect on the full-PSC optical properties, so that a rough FTO plate can be used as well as a smoother one as far as optical properties are concerned.

Finally, the percentage of incident light which is absorbed by the active layers constitutes the light-harvesting efficiency (LHE) and can be calculated from the simulated data of Fig. 4. Fig. 5a plots the LHE spectral dependence for the two (1D and 3D) models with roughness, along with the case without roughness. As expected, given the similarities of Figs 4(a) to (c), the three curves are very close. We compare these curves with the experimental external quantum efficiency (EQE) (ratio of the number of collected charges to the number of incident photons) of the solar cell (circles). The two curves are in overall qualitative agreement, however with a slight (up to 5-10 %) difference. If all absorbed photons lead to electron collection, the EQE and LHE values should be equal. The internal quantum efficiency (IQE), percentage of photon absorptions leading to electron collection, is equal to the ratio EQE/LHE and reflects losses in the charge transfer and collection processes. In figure 5(b) we give an estimate of the IQE by dividing the experimental EQE by the theoretical LHE. The values above unity (up to 1.05-1.1) are unphysical for the IQE because each absorbed photon cannot lead to more than one electron collection. This indicates that the calculated LHE is in fact slightly underestimated, most likely because MAPI absorption is higher than measured from ellipsometry or because the role of scattering in enhancing absorption is not fully described by the 3D model. This observation shows a limitation in attempting to model optically the PSC, but it is not unexpected, given the very complex optical nature of the device. However, it is satisfying to find that the IQE spectral dependence is, as expected, completely distinct from the spectral dependence of the device's optical properties contained in the LHE. This was not the case in previous reports on PSC modelling,[18, 30] which may illustrate the need for a precise full-device characterization in order to estimate the LHE accurately.

The IQE exhibits a continuous decrease by 10 to 20 % from 350 to 700 nm, probably because absorption in the red range occurs deeper in the MAPI layer (because MAPI absorbs less at these wavelengths) so that electron transfer to the TiO$_2$ ETL is less efficient. On figure 5(d), we estimate theoretically the average perovskite absorption position within the device, defined as $\langle zP_{abs}(x,y,z)\rangle / \langle P_{abs}(x,y,z)\rangle$ where the average is performed over the active layers. At shorter wavelengths, the absorption is located very close to the planar TiO$_2$ layer, so that the IQE is highest and probably close to unity. Light absorption then shifts deeper into the mesoporous TiO$_2$ layer and the MAPI layer as the wavelength increases, leading to IQE reduction.

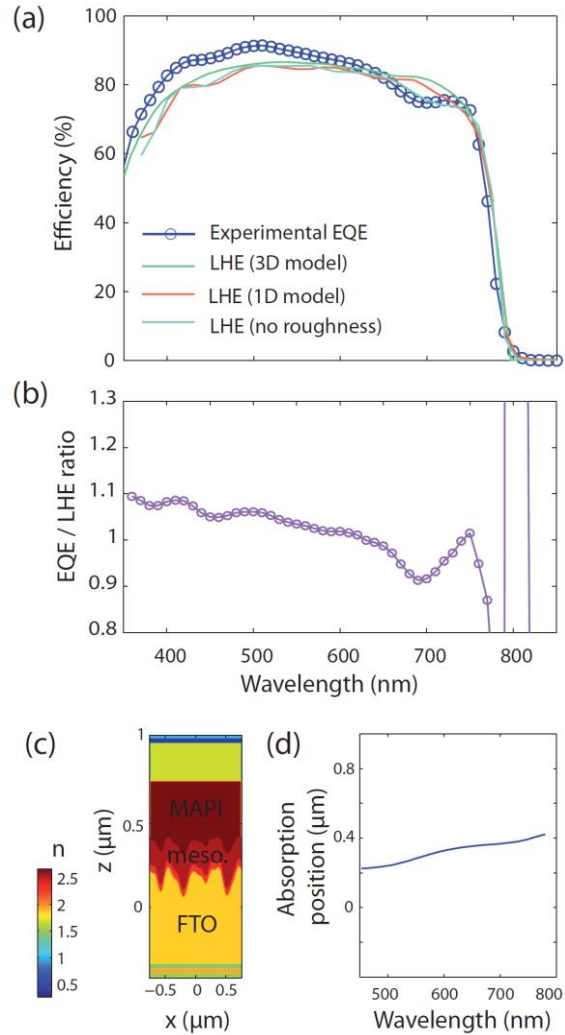

*Figure 5 : (a) Experimental EQE (blue dots) and theoretical LHE curves calculated from the 3D model (green) and the 1D model with (red) and without roughness (blue line). (b) Estimate of the internal quantum efficiency obtained by dividing the experimental EQE by the theoretical LHE (from the 3D model). (c) Index profile of the 3D model at 551 nm. (d) Theoretical light absorption position within the sample (from the 3D model), defined as $\langle zP_{abs}(x,y,z)\rangle / \langle P_{abs}(x,y,z)\rangle$, as a function of wavelength.*

Eventually, this work suggests several paths for light management improvement in such PSCs. (i) Over the MAPI absorption spectral range, a first limitation to the LHE is the reflection (around 4 %) of the incoming beam at the air-glass interface. This is a fundamental limit for a planar surface but it can be overcome by texturing the surface, as is already done in commercial silicon devices. For instance, a textured light-management foil added by nanoimprint lithography on the glass surface of a PSC has shown 5 % relative improvement of the cell performance.[38] (ii) A second optical limitation is absorption by the front electrode layers of $SnO_2$ and $SnO_2$:F. A thinner electrode may be used, provided that the charge conduction remains adequate. Indium-tin oxide (ITO) electrodes also seem a good alternative as lower evaluations of their parasitic absorption have been reported.[13, 19] (iii) A third, less important, optical loss mechanism is transmission through the gold electrode : it is negligible at 400 nm but increases up to 3 % at 700 nm and 9 % at 780 nm. A first way to avoid such losses would be to increase the MAPI layer's thickness, however at the risk of degrading the perovskite crystalline structure and reducing the charge transfer efficiency. A better alternative is to increase the thickness of the back-reflecting gold electrode. We model that an increase from 26 to 50 nm would reduce the device transmission in the visible range by a factor around 6, without any significant change in the gold and HTL absorption. Absorption by the active layer could then increase by an absolute 2 % at 700 nm, 3 % at 750 nm and 4 % at 780 nm. (iv) Parasitic absorption by the other materials (gold, spiro-OMeTAD, $TiO_2$) is negligible so that no improvement can be obtained from modifying these layers. (v) In terms of charge transfer within the device, the IQE decrease as a function of the wavelength can be attributed to charge generation deeper in the MAPI layer at higher wavelengths. This shows that the hole transfer undergoes minor losses while significant losses occur during electron transfer to $TiO_2$ (around 10-20 % at 700 nm). Increasing the thickness of the mesoporous $TiO_2$ layer could favor electron transfer without introducing optical losses.

**Conclusion :**

We have analyzed the effects of large substrate roughness (43 nm RMS) on the optical properties of a typical perovskite solar cell. We showed that ellipsometry is an effective technique for the complete characterization of PSC full layers stack. It provides a precise characterization in spite of the extreme complexity of the system. We tested the predictions of the resulting 1D model and found that, while a 3D model was necessary to describe the optical properties of the FTO plate alone, the full device could be described by the ellipsometric 1D model with little difference from the 3D model. Roughness effects within the full PSC device were shown to be quite limited, due to the low index contrast at the rough interfaces. We measured the PSC external quantum efficiency and found a good agreement of its spectral dependence with the calculated light-harvesting efficiency. This allowed us to probe the changes of the internal quantum efficiency over the absorption range and quantify the effect of the light penetration depth within the active layer on the electron transfer efficiency. Finally, we used the optical model to list the possible light-management strategies to improve the device.

**Methods :**

**Fabrication :** The FTO glass (TEC7, Pilkington) was purchased from Nippon Sheet Glass Co. Ltd. The compact $TiO_2$ layer was fabricated by spray pyrolysis method. The precursor solution of 7 mL isopropanol, 0.6 mL titanium isopropoxide, and 0.4 mL acetylacetone was sprayed onto substrates placed on a hot plate by a pressurized air gas gun and preheated at 455°C, 20 min before spray.[35,36]

The mesoporous $TiO_2$ layer was deposited by spin coating.[37] The $TiO_2$ paste (30-nm diameter spheres) was diluted in ethanol in a 1:8 mass ratio, dispersed by ultrasonic wave, then stirred for 12 h in brown bottles to avoid light. This solution was spin-coated at 5000 rpm (2500 rpm/s acceleration) for 20 s then dried at 125°C for 10 min. Afterward, the samples were annealed at 500°C for 30 min then let to cool down to around 200°C.

The perovskite layer was deposited by a one-step method based on the spin coating technique in a dry box.[35] The $CH_3NH_3PbI_3$ (MAPI) precursor solution consisted of 334 mg $PbI_2$, 115 mg methylammonium iodide (MAI), dissolved in 500 μL dimethyl sulfoxide (DMSO: $(CH_3)_2SO$). This precursor was spin-coated in two steps (1) at 1000 rpm (500 rpm/s acceleration) for 10 s and (2) at 6000 rpm (2000 rpm/s acceleration) for 30 s. Chlorobenzene anti-solvent of 100 μL was dropped 10 s before the end of the spin coating program to trigger the fast-crystallization of the perovskite material. The samples were then annealed at 105°C for 1 h.

The Spiro-OMeTAD layer was prepared by spin-coating as described in a previous work by some of us.[36] The solution was prepared by dissolving 72 mg Spiro-OMeTAD in 1 ml chlorobenzene. Then, 17.5 μL bis(trifluoromethylsulfonyl)imide lithium salt (LiTFSI) solution (520 mg in 1 mL acetonitrile (ACN)), 28 μL tert-butylpyridine (TBP) and 6 μL tris((2-1H-pyrazol-1-yl)-4-tert-butylpyridine)-cobalt(III)-tris(bis(trifluoromethylsulfonyl)imide) (300 mg in 1 mL of ACN) were added into this solution, 35 μL of which was spin-coated at 4000 rpm (2000 rpm/s acceleration) for 30 s.[36]

Finally, a thin layer of gold was thermally evaporated through a mask to act as an electrode and back reflecting mirror.

**Ellipsometric characterization :** Ellipsometric measurements were performed by a V-VASE ellipsometer from J.A. Woollam Co., Inc. A translucent tape was applied to the bottom of the sample to eliminate back reflection, which might interfere with the surface reflection, and a back reflection correction parameter was introduced. The scanning wavelength range was from 300 to 2000 nm in steps of 10 nm and incidence angles of 50°/60°/70°. The complex reflectance ratio ρ is defined as $\rho = \frac{r_p}{r_s} = \tan\Psi \cdot e^{i\Delta}$, where $\tan\Psi$ is the amplitude ratio and Δ is the phase difference between the two p and s polarizations. Then the software WVASE32 was utilized to fit a model of stacked layers of appropriate optical constants and thicknesses to the experimental curves $\Psi(\lambda)$ and $\Delta(\lambda)$.

**Transmission spectra :** Transmission spectroscopy was carried out by Cary 5000 UV-Vis-NIR spectrophotometer. Our measurement range of wavelength was from 300 nm to 2000 nm with a spectral resolution of 1 nm.

**3D simulations :** The simulations were performed using the FDTD method with Lumerical software. We defined a 3D-computing cell (1.5μmx1.5μm in x and y the dimensions of the layers and 8.5μm in z, the dimension of the stack) with periodic conditions in x and y and phase-matching layers to avoid parasitic reflections in z.

The stack (z= -0.5 μm to z= 0.9μm), deposited on a 6-μm microscope glass slide (from z= -6.5μm to z= -0.5μm), was illuminated by a plane wave at normal incidence, located at z= -5.5μm (inside the glass slide to avoid interference fringes due to the air/glass and stack/air interfaces). The transmission and reflection spectra were corrected by the transmission of the air/glass interface to be compared to the experimental ones.

The light source had a spectral range from 370 to 2000 nm for transmission and reflection simulations (resolution: 8.15 nm) and from 450 to 1200 nm for absorption simulations (resolution : 7.5 nm). For reflection and transmission simulations, we recorded the transmitted power, normalized by the source incident power, on two planes located at z= -6.5μm for reflection and z= 1.8μm for transmission.

For absorption simulations, we used a 3D-detector to record the electric field amplitude and the real and imaginary parts of the optical index at different points (x, y z) for 100 values of the wavelength (from 450 to 1200 nm). The resolution was 20 nm in x and y and 6.5 nm in z.

The roughness of the FTO and $TiO_2$ layers were simulated using the random surface tool of Lumerical. The roughness is generated by creating a random matrix of values in k space defined by the spans in x and y. A Gaussian filter is applied to this matrix, then a Fourier transform is used to transform the matrix back to real space. The roughness used was characterized by an RMS amplitude of 50 nm (for the FTO and $TiO_2$ compact layers) and of 30 nm (for the $TiO_2$ mesoporous layer) and by a correlation length of 100 nm in x and y.

**Acknowledgements**

Mr D. Z. acknowledges the CSC-Paristech program for scholarship funding (grant number 201806310126). Prof. T. P. acknowledges the Agence Nationale de la Recheche (ANR) for financial support via the Moreless project ANR-18-CE05-0026.

# Conflicts of interest

There are no conflicts to declare.